\documentclass[12pt]{article}

\usepackage[pdftex]{graphicx}

\begin{document}
\font\ninerm = cmr9
\baselineskip 12pt plus .5pt minus .5pt

\def\footnoterule{\kern-3pt \hrule width \hsize \kern2.5pt}

\pagestyle{empty}


\begin{center}
{\large\bf Measurement of the space-time interval
between two events\\
using the retarded and advanced times of each event\\
with respect to a time-like world-line\\}
\end{center}
\vskip 1.5 cm
\begin{center}
{\bf Giovanni AMELINO-CAMELIA}$^a$ and {\bf John STACHEL}$^b$\\
\end{center}
\begin{center}
{\it $^a$Dipartimento di Fisica,
Universit\'{a} di Roma ``La Sapienza",\\
and INFN, Sezione Roma1, P.le Moro 2, Rome, Italy$~~~~$}\\
{\it $^b$Department of Physics and Center for Einstein Studies,$~~~~$\\
Boston University, Boston, MA 02215, U.S.A.$~~~~~~~~~~~~~~$}
\end{center}

\vspace{1cm}
\begin{center}
{\bf ABSTRACT}
\end{center}

{\leftskip=0.6in \rightskip=0.6in
Several recent studies have been
devoted to investigating the limitations that ordinary quantum
mechanics and/or quantum gravity might impose on the measurability
of space-time observables. These analyses are often confined to
the simplified context of two-dimensional flat space-time and rely
on a simple procedure for the measurement of space-like distances
based on the exchange of light signals. We present a
generalization of this measurement procedure applicable to all
three types of space-time intervals between two events in
space-times of any number of dimensions. We also present some
preliminary observations on an alternative measurement procedure
that can be applied taking into account the gravitational field of
the measuring apparatus, and briefly discuss quantum limitations
of measurability in this context. }

\newpage

\baselineskip 14pt plus .5pt minus .5pt

\pagenumbering{arabic}
\pagestyle{plain}

\section{Introduction}
The limitations on measurability encountered in ordinary
(non-gravitational) applications of quantum mechanics concern
conjoint measurement of pairs of noncommuting observables. In particular,
any given observable can be measured with arbitrary
accuracy at the cost of renouncing any attempt at measurement of a conjugate
observable.
However, a number of arguments have been offered in support
of the possibility that gravitational observables, including both
the chrono-geometric and inertio-gravitational structures, may
be subject to more severe measurability limitations. In particular,
there may be absolute limitations (i.e. not avoidable even by sacrificing
all knowledge of their conjugates) on the measurability of certain of these observables.
These results have been obtained within the framework of certain approaches to quantum
gravity, most notably in string theory~\cite{venegross}, and
they are also supported by certain heuristic
analyses~\cite{mead,padma,dopl1994,ahlu1994,ng1994,gacmpla,garay},
in which the gravitational degrees of freedom playing a role in
the relevant measurement procedures are treated classically,
while the non-gravitational degrees of freedom are
treated using ordinary quantum field theory in flat or non-flat background space-times.
In addition, there are good heuristic arguments (see, {\it e.g.},
Ref.~\cite{dowkerNS180e36})
essentially going back to Bronstein~\cite{bronstein},
that the very concept of a smooth space-time should break down
at the Planck four-volume. An attempt to localize a massive object
to within a volume approaching the Planck volume
will give rise, according to the Heisenberg indeterminacy relations,
to energy fluctuations that will produce a black hole lasting a Planck time,
making any further space-time localization meaningless (see, e.g., Ref.~\cite{thiemA}, p.~117).
These arguments suggest that space-time smoothness breaks down at
the Planck scale; and the same conclusion is reached by exploring
heuristically the possible consequences of quantization of the gravitation field
(see, e.g., Ref.~\cite{thiemB}, p.~40).

The main interest of these results lies in the hope
that they might provide key insights concerning the structure
of a future theory
encompassing both present-day quantum
field theory and general relativity (``quantum gravity").
Of course, faith in the reliability of these hints depends
not only on the rigour of the individual analyses, but also on
the generality (or representative nature) of the measurement
procedures analyzed in establishing the suggested limitations.
We share the point of view
emphasized by Heisenberg~\cite{heise} and Bohr and Rosenfeld~\cite{rose},
that the limits of {\it definability} of a quantity within any formalism should
coincide with the limits of {\it measurability} of that quantity
for all conceivable (ideal) measurement procedures.
For well-established theories, this criterion can be
tested. For example, in spite of
a serious challenge~\cite{lanper},
source-free quantum electrodynamics was shown
to pass this test~\cite{rose}.
In the case of quantum gravity, our situation is rather the opposite.
In the absence of a fully accepted, rigorous theory,
exploration of the limits of measurability
of various quantities can serve as a tool to provide clues
in the search for such a theory:
If we are fairly certain of the results of our measurability analysis, the proposed
theory must be fully consistent with these results.

This paper is concerned with the generalization of one of the most
frequently-cited procedures in such discussions of limitations on
quantum measurement of geometrical quantities, first discussed in
detail by  Wigner and Salecker~\cite{wign,salWig}: the measurement
of space-like intervals in two-dimensional space-time based on the
exchange of light signals.

Analyses based on this measurement procedure have been
used extensively (see, {\it e.g.}, Refs.~\cite{ng1994,gacmpla,adlerSW,gacgwi})
as the source of intuitive ideas about the quantum-gravity problem.
In evaluating the reliability of such intuitions, one
might  be concerned by the fact that
their starting point only considered
space-like distances in two dimensions; this
might appear to be too narrow a basis on which to establish
general conclusions.
Another reason for possible concern is the way in
which Wigner-Salecker handled the delicate issue of the relation
between the microscopic aspects of quantum phenomena and
the macroscopic devices used in recording measurements of them.

This paper provides a simple
generalization of the Wigner-Salecker procedure
to the case of a space-time interval of any type
defined by two events; and, perhaps more significantly,
in any number of space-time dimensions\footnote{For Wigner's
concerns with the limitation to two dimensions,
see Ref.~\cite{wign}, pages 261 and 262.}.
We also propose a strategy for development of an alternative measurement
procedure that handles more carefully the relation between
its microscopic and macroscopic aspects.

We could turn directly to the case of
arbitrary space-time intervals in a space-time of an
arbitrary number of dimensions.
But we prefer to adopt a more
pedagogical approach, proceeding inductively from the case of
2D (two-dimensional) space-time,
to 3D and then to (4+n)D
space-times, where n=0,1,2....

The paper is organized as follows.
Section~2 provides a brief summary
of the Wigner and Salecker measurement analysis
for space-like intervals in 2D Minkowski space-time.
Section~3 is still confined to the 2D case, but generalizes the result to
space-time intervals of any type.
Section~4 concerns 3D Minkowski space-time
with emphasis on some new elements required for
measurability analyses in more than two dimensions.
The straightforward generalization
to (4+n)D space-times is discussed briefly in Section~5.
In Section~6 we consider some limitations of the
Wigner-Salecker analysis that result from the nature of
the clock they employ
(microscopic versus macroscopic), and the way in
which the microscopic aspects of the measurement process
are eventually recorded by a macroscopic device.
We argue that the Wigner-Salecker procedure
can be improved by letting the macroscopic recording
apparatus consist of a spherically symmetric shell, entirely surrounding the region under study.
In Section~7 we observe that this arrangement still can be used when
the gravitational field of the surrounding shell is taken into account,
and briefly consider some implications for the quantum-gravitational measurement problem.
Section~8 summarizes our findings and indicates some goals
of our future work.

\section{Summary of the SW analysis (2D, space-like interval)}
Rather than following Wigner-Salecker's original measurement
procedure for 2D space-like intervals in detail, we present the relevant portion
of their analysis
in a language and notation that provide a better starting
point for our generalization, emphasizing
those aspects of their work that have
been cited and used most in recent studies~\cite{ng1994,gacmpla,adlerSW}.

Wigner and Salecker~\cite{wign,salWig}
determine the space-like interval between two events in terms of certain
proper time
intervals\footnote{Note that, like  Wigner and Salecker, we
take for granted that the time interval between two time-like separated events
can be directly measured by the proper time readings
of a clock moving on the time-like geodesic
between the two events. At the quantum level this assumption might call for
further analysis.}
along a given time-like world line passing through one of the events:
namely, the time intervals for
massless probes (photons) to traverse the
distance back and forth between the world line and the two
 events\footnote{Wigner and
Salecker also provide~\cite{wign,salWig}
a critique of the use of rigid rods in the measurement
of distances. For a defense of legitimacy of their use
see Ref.~\cite{stachel983}.}.
The crucial ingredient is the relation between
these proper time intervals
and the space-like interval,
so we start by expressing a space-like interval
between two events in 2D Minkowski space-time
in terms of proper times on an inertial time-like world-line PQ (see Figure 1).
The simplest case is when one of the two
events lies on PQ and the interval between the two events
is orthogonal to PQ.
As we shall see in Section 3, the interval between
any two events can be derived
easily from this case.

\begin{figure}
\centering
\includegraphics[height=120mm]{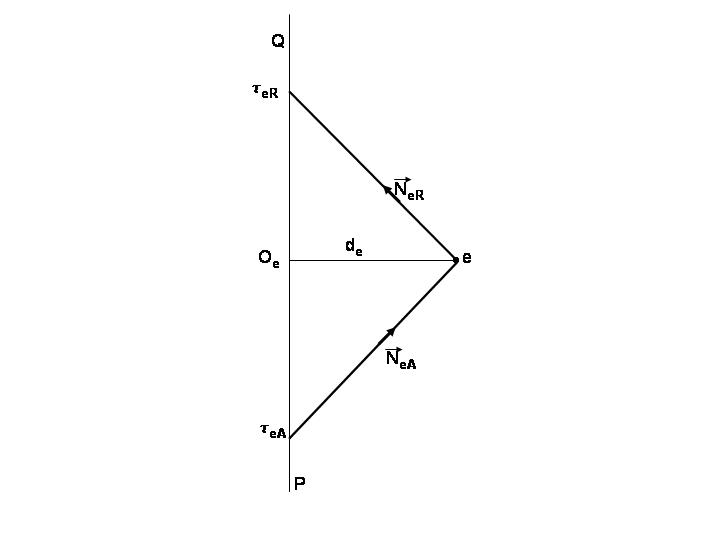}
\caption{Proper-time measurements on the inertial time-like world-line PQ
used to measure the length $d_e$ of the space-like interval between the event $e$
and the world line.}
\label{fig1}
\end{figure}

Let $\tau_{e,A}$ be the (proper)
time of the point on PQ such that a light signal
sent from it reaches the event $e$,
and $\tau_{e,R}$ the (proper) time a signal sent from $e$ will reach
the world-line. (We shall sometimes refer to these as
the ``advanced" and ``retarded"
times of the event $e$ with respect to PQ.)
Letting
\begin{equation}
\Delta \tau_e \equiv  \tau_{e,R} - \tau_{e,A}
\label{deltataue}
\end{equation}
and calling $\vec{\Delta \tau_e}$
the corresponding (2D) space-time vector (in general, $\vec{V}$ will denote
a space-time vector of magnitude $V$),
we see that the interval (distance) $\vec{d_e}$
between the event $O_e$ at the midpoint of $\vec{\Delta \tau_e}$
(corresponding to proper
time $\tau_e = \tau_{e,A}/2 + \tau_{e,R}/2$)
and the event $e$
can be found by ``squaring" the vectorial equation\footnote{We use units with $c=1$.}:
\begin{equation}
\vec{d_e} + \vec{N_{e,R}}
= \vec{\Delta \tau_e}/2
\label{eqde}
\end{equation}
where $\vec{N_{e,R}}$ is the retarded null
vector connecting $e$ and the event on the PQ world-line
that corresponds to the proper time $\tau_{e,R}$.

 Using the indefinite Minkowski
metric to ``square (\ref{eqde})", and observing that by construction $\vec{d_e}$
is space-like while  $\vec{\Delta \tau_e}$
is time-like and that the square of a null vector $=0$,
we get
\begin{equation}
d_e = \Delta \tau_e/2 = (\tau_{e,R} - \tau_{e,A})/2
\label{solde2d}
\end{equation}
Thus, one can determine $d_e$, the space-like distance between the
event $e$ off the world-line PQ and the event $O_e$ on PQ,
by measuring the advanced and retarded
times $\tau_{e,R}$ and $\tau_{e,A}$.

A comment is important for our generalization: In 3D Minkowski space,
we only need the half-plane
defined by $PQ$ and $e$ since one can
generate three-dimensional Minkowski space
by rotating this half-plane about the line PQ.
Put another way, the spatial direction orthogonal to the
world-line can be interpreted as a radial direction.
In their 2D diagrams, Salecker and Wigner
correctly picture the entire two-dimensional Minkowski plane,
but, in calculating the distance between two events, they tacitly
assume that both lie in the same half-plane, which need not always be the case.
This observation is important
because they assert~\cite{wign,salWig} that, confining themselves to two dimensions,
they are able\footnote{This also tacitly assumes the availability of
a totally-reflecting mirror ({\it i.e.}, with zero transmission coefficient).}
to avoid the problem of ascertaining the direction from which the
light signals come when calculating the interval between two events.
This is not quite correct, as can be seen by considering how one
might distinguish between two events that have the {\it same} advanced
and retarded times with respect to the world-line but lie in {\it opposite}
half-planes. This is the remnant in two dimensions of the fact that,
in the higher-dimensional cases, we cannot avoid the
problem of ascertaining the angle between
the directions of two light signals
(rotating a half-plane through the angle $\pi$ generates the other half-plane).

\section{Arbitrary interval in 2D space-time}
As we have seen, the Wigner-Salecker analysis is satisfactory,
and even enlightening.
However, it does have definite limitations, notably
 the restrictions
to space-like intervals and 2D space-times.
In this section, we remain in 2D Minkowski space-time
but generalize the
Wigner-Salecker analysis
to arbitary
space-time intervals defined by two
events that may both lie off the time-like world-line PQ.

Our measurement procedure is based on
the calculation of the space-time interval defined
by any two events off PQ
in terms of proper time measurements on PQ
Let $e$ and $u$ be the two events, $d_{eu}$
the space-time interval between them,
and $\vec{d_{eu}}$ the corresponding (2D)
space-time vector.
In the same way
that $O_e$ was defined with respect to $e$ in Section 2,
one can define
an event $O_u$  on the world-line with respect to $u$,
and a vector $\vec{d_{u}}$
connecting $O_u$ and $u$.
It is then easy to see from Figure 2 that
\begin{equation}
\vec{d}_{eu} =
\vec{D}_{eu} + \vec{d}_{u} - \vec{d_{e}} ~,
\label{eqdeu2d}
\end{equation}
where $\vec{D_{eu}}$ is the vector connecting
the events $O_e$ and $O_u$ on the world-line PQ.
Note that by construction $\vec{D_{eu}}$ is time-like and
$\vec{d_{u}}$ and $\vec{d_{e}}$
are space-like,
while $\vec{d_{eu}}$ can be any type of
vector ({\it e.g.}, time-like
if $\vec{d_{u}} = \vec{d_{e}}$
or space-like if $\vec{D_{eu}} = 0$).

\begin{figure}
\centering
\includegraphics[height=120mm]{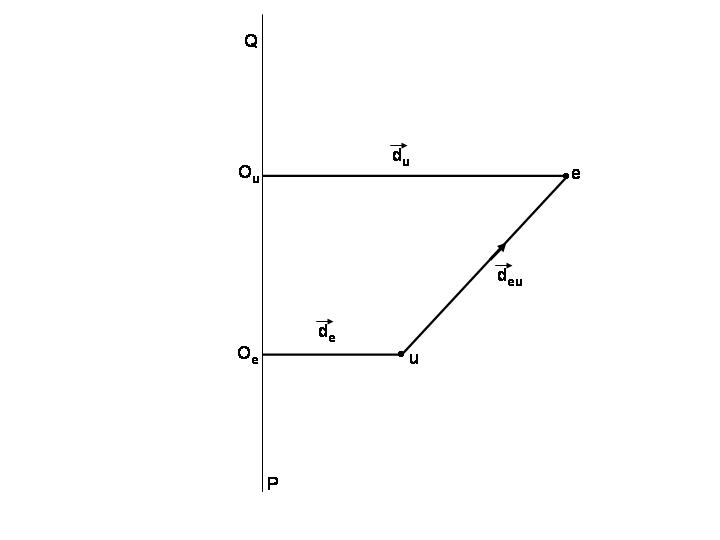}
\caption{The space-time interval $\vec{d_{eu}}$ defined
by the events $e$ and $u$
in terms of proper time measurements on the world-line PQ. $\vec{d_{e}}$
and $\vec{d_{u}}$ are the spatial intervals between the world-line PQ and $e$ and $u$,
respectively.}
\label{fig2}
\end{figure}

Using simple generalizations of the notation and
arguments used in Section 2, one easily finds
that $d_e = (\tau_{e,R} - \tau_{e,A})/2$,
$d_u = (\tau_{u,R} - \tau_{u,A})/2$,
and $D_{eu} = (\tau_{e,R} + \tau_{e,A}
- \tau_{u,R} - \tau_{u,A})/2$.
It is then straightforward to show that (adopting Minkowski signature $+,-,-,-$)
\begin{eqnarray}
\vec{d_{eu}} {\cdot} \vec{d_{eu}}
& = & D_{eu}^2 - |d_u -d_e|^2 = \cr
& = & {1 \over 4} [(\tau_{e,R} + \tau_{e,A}
- \tau_{u,R} - \tau_{u,A})^2 -
(\tau_{u,R} - \tau_{u,A} - \tau_{e,R} + \tau_{e,A})^2] \cr
& = & \tau_{e,R} \tau_{e,A}
- \tau_{e,A} \tau_{u,R} - \tau_{e,R} \tau_{u,A}
+ \tau_{u,R} \tau_{u,A}
~,
\label{soldeu2d}
\end{eqnarray}
which indeed
expresses $\vec{d_{eu}} {\cdot} \vec{d_{eu}}$
in terms of the (four) advanced and retarded times of the
(two) events $e$ and $u$.

As mentioned in Section 1, we postpone to a later paper
a detailed analysis of the implications of this result
for measurements at the quantum level.
It is however reassuring to note that our generalization
to arbitrary intervals
does not require any qualitatively new type of measurement.
All required information can still be obtained by recording some
(advanced and retarded) readings of a clock placed close
to the reference world-line.
It should therefore be straightforward to generalize to
arbitrary intervals the arguments previously advanced~\cite{ng1994,gacmpla,adlerSW}
in analyses of the implications
 at the quantum level of the measurement of space-like intervals
originally considered by Wigner-Salecker.

\section{Arbitrary interval in 3D space-time}
As noted in Section 2, we can generate 3D Minkowski space-time from the 2D
space-time of the previous two Sections
by rotating the
half-plane about the time-like world-line PQ
through the angle $2 \pi$.
For any given event $e$,
one can still define
the retarded and advanced null
vectors $\vec{N_{e,R}}$ and $\vec{N_{e,A}}$ that,
after the $2 \pi$ rotation, lie respectively
on the retarded and advanced null cones having their origin
on PQ and including the point $e$.
These retarded and advanced null cones intersect in a circle that lies in
the space-like plane bisecting
the interval vector $\vec{\Delta \tau_e}$;
the center of the circle is at the
bisecting point $O_e$ of the world-line and $e$ of course
lies on this circle.

Before discussing the most general space-time interval
defined by any two events $e$ and $u$ off PQ,
let us first consider the special case, in which $e$
and $u$ are such that the points $O_e$ and $O_u$ coincide (see Figure 3),
{\it i.e.} both $e$
and $u$ lie in the same space-like plane intersecting
the world line PQ in a single point $O_e=O_u$.
In this case, the distance $\vec{d_{eu}}$ is
space-like and may be found from the space-like triangle
with sides $\vec{d_{eu}}$, $\vec{d_{e}}$ and $\vec{d_{u}}$:
\begin{equation}
\vec{d_{eu}} = \vec{d_{u}} - \vec{d_{e}}
~.
\label{eqdeu3dsimple}
\end{equation}
 ``Squaring" this (which amounts to rederiving the law of cosines),
we get:
\begin{equation}
\vec{d_{eu}} {\cdot} \vec{d_{eu}}
= - |\vec{d_{u}} - \vec{d_{e}}|^2 =
- |d_{e}|^2
- |d_{u}|^2
+ 2 |d_{e}| |d_{u}| \, cos(\phi)
~,
\label{soldeu3dsimple}
\end{equation}
where $d_{e}$ and $d_{u}$ are given respectively by
$d_e = (\tau_{e,R} - \tau_{e,A})/2$ and
$d_u = (\tau_{u,R} - \tau_{u,A})/2$,
and $\phi$ is the angle (in the plane on which both lie)
between $\vec{d_{e}}$ and $\vec{d_{u}}$;
{\it i.e.}, the angle between two radii of the circle, discussed above,
defined by the intersection of retarded and advanced null cones.

\begin{figure}
\centering
\includegraphics[height=120mm]{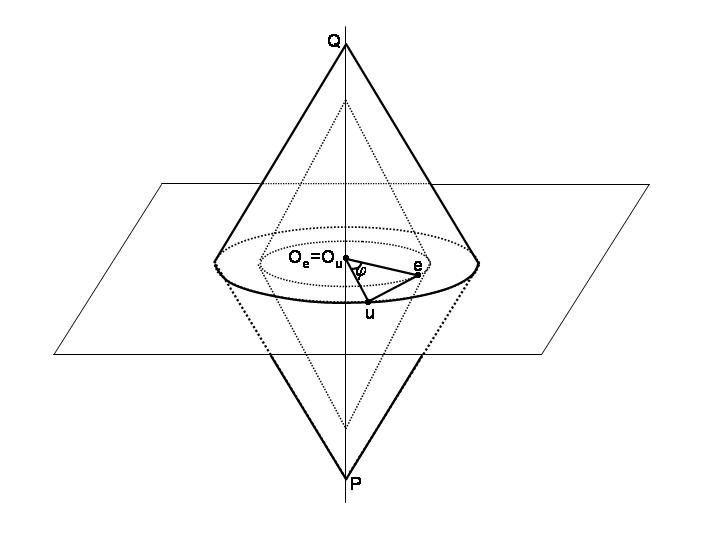}
\caption{The special case in which $e$
and $u$ are such that the points $O_e$ and $O_u$ coincide.}
\label{fig3}
\end{figure}

It is not hard now to generalize formula (\ref{soldeu3dsimple})
to the case of arbitrary $e$ and $u$: In general $O_e$ and $O_u$ need
not coincide ({\it i.e.}, $e$ and $u$ do not generally
lie in the same space-like plane).
In this case, $\vec{d_{eu}}$
can be a  vector of any type,
but of course $\vec{d_{e}}$ and $\vec{d_{u}}$
are (by construction) still space-like.
As in Section~3, it is also convenient here
to introduce the time-like
vector $\vec{D_{eu}}$ connecting
the points $O_e$ and $O_u$ on the world-line PQ,
allowing us to generalize Eq.~(\ref{eqdeu2d}) and (\ref{eqdeu3dsimple}) for $\vec{d_{eu}}$:
\begin{equation}
\vec{d_{eu}}
= \vec{D_{eu}} + \vec{d_{u}} - \vec{d_{e}}
~.
\label{eqdeu3d}
\end{equation}
Squaring this, we obtain
\begin{equation}
\vec{d_{eu}} {\cdot} \vec{d_{eu}}
= D_{eu}^2 - |\vec{d_{u}} - \vec{d_{e}}|^2 =
D_{eu}^2 - |d_{e}|^2
- |d_{u}|^2
+ 2 |d_{e}| |d_{u}| \, cos(\phi)
~,
\label{soldeu3d}
\end{equation}
where $\phi$ is still the angle\footnote{More precisely,
taking into account that
here $\vec{d_{u}}$ and $\vec{d_{e}}$
do not lie in the same plane, $\phi$ is the angle between $\vec{d_{u}}$
and the translation (which is actually a translation
by $\vec{D_{eu}}$) of $\vec{d_{e}}$
to the plane, in which $\vec{d_{u}}$ lies.}
between $\vec{d_{e}}$ and $\vec{d_{u}}$.

Using (\ref{soldeu3d}), we can
express $\vec{d_{eu}} {\cdot} \vec{d_{eu}}$
in terms of advanced and retarded times on the world-line PQ:\footnote{We thank
Andor Frenkel for pointing out to us the possibility
of simplifying Eq.~(\ref{soldeu3dfinal}) to its final form.}
\begin{eqnarray}
\vec{d_{eu}} {\cdot} \vec{d_{eu}} \!\! &=& \!\! {1 \over 4}
[(\tau_{e,R} + \tau_{e,A} - \tau_{u,R} - \tau_{u,A})^2 -
(\tau_{u,R} - \tau_{u,A})^2 - (\tau_{e,R} - \tau_{e,A})^2
\nonumber\\
& & ~~- 2 |\tau_{u,R} - \tau_{u,A}| |\tau_{e,R} - \tau_{e,A}|
\cos(\phi)]
\label{soldeu3dfinal}
\\
&=& \tau_{e,R} \tau_{e,A} + \tau_{u,R} \tau_{u,A}
- \cos^2 (\phi) [\tau_{e,A} \tau_{u,R} + \tau_{e,R} \tau_{u,A}]
- \sin^2 (\phi) [\tau_{e,A} \tau_{u,A} + \tau_{e,R} \tau_{u,R}]
~. \nonumber
\end{eqnarray}

Anticipating the quantum measurability
analysis, we observe that
the most significant new element that has emerged
in generalizing the Wigner-Salecker procedure to 3D space-times
is the requirement that the angle $\phi$ also be measured.
All other measurements still involve only
(advanced and retarded) readings of a clock placed close
to the reference world-line. Interestingly (if perhaps not surprisingly),
the way to minimize the sensitivity of the interval measurement to a possible
uncertainty in the angle $\phi$ is by means of
an arrangement making $\phi \sim 0$ (or $\phi \sim \pi$),
{\it i.e.} an arrangement in which the experimenter is on a world line such
that $\vec{d_{e}}$ and $\vec{d_{u}}$ are parallel (or anti-parallel).

\section{Generalization to 4D space-time}
Having obtained the formulas for the 3D case,
the generalization to 4D Minkowski space-time
is now rather trivial.
Instead of rotation through a circle, to get four-dimensional Minkowski space-time,
we need merely rotate the 2D Minkowski
half-plane through a two-sphere of spherical angle $4 \pi$
around the line PQ to get four-dimensional Minkowski space-time.
Instead of a circle, the advanced and retarded light
cones with vertices on PQ and
including the event $e$ now intersect in a two-sphere
lying in a space-like
hyperplane ({\it i.e.}, a three-space) orthogonal to PQ and
centered on the midpoint $O_e$
of the interval $\vec{\Delta \tau_e}$.
The geometrical constructions then proceed just as
in the 3D case, but now, instead of circles, $e$ and $u$ lie on spheres centered
on the world-line (instead of circles).
One can again introduce the space-like
vectors $\vec{d_{e}}$ and $\vec{d_{u}}$
and the time-like vector $\vec{D_{eu}}$
in complete analogy with the 2D and 3D cases.
In case both events $e$ and $u$ lie
in the same space-like hypersurface
orthogonal to PQ, the two spheres
are concentric; and in order to use Eq.~(\ref{soldeu3dsimple})
in 4D, we merely need to
interpret $\phi$ as the angle
between two radii of the sphere.

Similarly, equations (\ref{soldeu3d}) and (\ref{soldeu3dfinal})
are still valid, and, just as it is in the 3D case, the
distance $\sqrt{\vec{d_{eu}} {\cdot} \vec{d_{eu}}}$
is determined by the (four) advanced and retarded
times of the (two) events with respect to
the world-line and by the angle $\phi$.

Should the need arise for generalization to higher dimensions,
it is obvious how to proceed.

\section{Beyond Wigner-Salecker: why introduce the microscopic clock? }
The analysis carried out in the previous Sections
eliminates several of the limitations of the
original Wigner-Salecker analysis: rather than being restricted
to the context of space-like intervals in 2D space-time,
the measurement procedure can now be applied to arbitrary
space-time intervals in a space-time of Minkowski signature
and arbitrary dimension, using an arbitrary\footnote{The
original Wigner-Salecker analysis assumed that
one of the two events defining the interval
was on the reference world-line.}
time-like world-line.
So far we have only alluded to some other limitations of the Wigner-Salecker
analysis relevant to the key issue
of the transfer of information about the microscopic system
under observation to the macroscopic devices used to ultimately
record the measurement result. This information is needed to complete the process for which we
must compute a probability.
If the process is treated
classically, this probability
can be computed directly from the ensemble in phase space defined by the process.
If the process is treated quantum mechanically, a probability amplitude
for it must be computed.\footnote{We use Feynman's term ``process" to
describe what Bohr calls a ``phenomenon": The preparation by some external apparatus
of a quantum system, which then undergoes some interaction(s), the result of which is
then ``registered" by another external apparatus. The aim of any quantum-mechanical
formalism is the computation of a probability amplitude for any such process
(see Ref.~\cite{stach1997}).For a discussion of the relations between classical and
quantum ensembles, see Ref.~\cite{stachel2005}.}

The first limitation of this type was recognized by Wigner and Salecker.
Early in their discussion of the microscopic clock,
they remark (p.~571 of Ref.~\cite{salWig}):

\begin{quotation}
{\it As is well known,
and as was pointed out most clearly by von Neumann,
the measurement is not completed until its result is recorded by some
macroscopic object. If the macroscopic object were part of the clock,
no microscopic clock could exist. The way out of this difficulty is to
transmit the signal of the clock to a macroscopic recorder (which can
be the `final observer') which is far away from the clock,
considered from the point of view of the average motion of the latter.
The transmitting signal will be considered part of the [microscopic]
clock, not, however, the recording apparatus.}
 \end{quotation}

This is essentially the reason why they confined
themselves to a world of one spatial dimension (pp.~571-572 of Ref.~\cite{salWig}):

\begin{quotation}
{\it If the
transmitting signal is to be microscopic,
that is, if it is to consist of only a few quanta
(actually, our signals will be light quanta), it will
reach the recording equipment with certainty only if it
does not spread out in every direction. In order to
guarantee this, we confine ourselves to a world which
has, in addition to the time-like dimension, only one
space-like dimension.}
\end{quotation}

\noindent
They do not discuss any further the nature of the distant macroscopic
recording device.

Another, related limitation
was not pointed out by Wigner and  Salecker. They take as
unproblematic the notion of inertial paths, and indeed
parallel inertial paths, in space-time
(see Fig. 3, p. 573, Ref.~\cite{salWig}, for example); as well as
the notion of an inertial frame of reference
(see Fig. 1, p. 572, Ref.~\cite{salWig}, for example).
The relation between these two
limitations is that the macroscopic
recording device can also serve to fix the
inertial frame of reference, or at least be rigidly
related to the macroscopic system that serves this purpose.

The relation between the microscopic clock, including the
light quanta that transmit the signal, and the macroscopic
recording device serving to fix the frame of reference, is
quite analogous to that described by Bohr when he considers
two possible alternative one-slit diaphragm experiments. In one case, the
diaphragm is rigidly attached to the apparatus defining
the inertial reference frame; and in the other, it is suspended
from the frame by springs (see pp.~697-698 of Ref.~\cite{bohrPAGES1}
and pp. 218-221, esp. Figs. 4, p. 219 and Fig. 5, p. 220,
of Ref.~\cite{bohrPAGES2}).
Until the relation of the diaphragm to the macroscopic
apparatus defining the inertial frame of reference is
fixed, one cannot speak of a definite phenomenon or process.
The rigidly-fixed diaphragm can only be used for
position measurements; while the suspended diaphragm can also be
used for momentum measurement (for a fuller discussion,
see Refs.~\cite{bai,baistachel}).

Applying Bohr's point of view to the case of the (microscopic) clock,
until the macroscopic recorder has been introduced,
serving both to fix or define an inertial frame
of reference and to register the outcome of the process,
one cannot meaningfully discuss
the motion of the clock with respect to an inertial frame.
In the context of special relativity, no more need
be added, since the points of space and the instants of
time relative to the inertial frame  (assuming
the Poincar\'e-Einstein convention to define the global time of the inertial frame)
can be individuated
quite independently of the quantum physical processes under investigation.
This is not the case in general relativity, the theory in which we are really interested.
But before turning to that case,
let us return to the problem (discussed in Section 2)
that limited Wigner and Salecker to one
spatial-dimension: the spreading of the light signal.
We can get around that problem by imagining
the macroscopic recording apparatus to entirely surround the
region under study\footnote{This actually
implements Wigner's suggestion, at page 263 of
Ref.~\cite{wign}: {\it ``In our experiments we surround the microscopic
objects with a very macroscopic framework and observe coincidences
between the particles emanating from the microscopic system,
and parts of the framework"}} -- for example, to consist of a spherical shell
of matter,
within which lies the clock, as well as all the events to be investigated
with it. Then, a record of the places and the times
on the shell, at which the signals from the clock are
received, may be treated as final observer results in von Neumann's
terminology (see above) or just the results of the measurement
defining the process or
phenomenon under investigation in the Bohr-Feynman terminology.
As we shall show, these data then can be used to ascertain the
macroscopic clock times
needed to define the space-time interval between two events.

\section{Beyond Wigner-Salecker: why not introduce gravity?}
So far,
the discussion has entirely neglected gravity and hence has
been confined to special relativity. However, the
introduction of a massive spherical
shell offers an obvious way to begin to remedy this defect.
Inside a spherical ``hole" (matter- and non-gravitational
field-free region of space-time) within a spherically-symmetric
distribution of matter,  both hole and shell having the same center of symmetry,
the Minkowski metric is a solution to the field equations of general
relativity. So it appears that we can, to a certain extent, have our
cake and eat it: We can carry certain results of a special-relativistic
analysis into a valid general-relativistic context.

But that is the case only to a certain extent.
Due to the universality of gravity, the presence within the shell
of a clock, light signals,
and whatever physical processes are used to define the events,
the interval between which we wish to measure,
will destroy the spherical symmetry. We must therefore assume that the masses
(or more properly the physical components of the stress-energy tensors)
of all such entities
are so small compared to the mass of the shell that,
to  a good first approximation, we may neglect their effect on the
gravitational field inside the shell\footnote{One feature of the general-relativistic
version of the material shell model must be noted here,
even though we postpone detailed consideration of its
implications until a later paper. In order that such a static shell be possible in
general relativity, we must ensure that the radius of the shell falls
outside the Schwarzschild radius associated with the mass of the shell
(see the Introduction).
That is, we must assure that $2GM/c^2R$ be less than unity. This means,
of course, that if for any reason we are forced to increase the mass
of the shell (for example, in order that we may treat objects
inside the shell of mass of the order $m$ as test bodies to a better
and better approximation, we shall be forced to make the ratio $m/M$
smaller and smaller), we must take care to increase the radius $R$ if we start to
approach the Schwarzschild limit.}. To the next approximation,
we might use the linearized Einstein equations to take account of
modifications of the gravitational field inside the shell, but
we postpone study of this question until a later paper.\footnote{For a careful
quantum measurability analysis
of the linearized gravitational field,
see Ref.~\cite{bergstac} (also see Ref.~\cite{stachelY}).}

Perhaps an even more important modification in the general-relativistic
case is the loss of a field-independent definition of spatial points and instants
of time in the empty region within the shell.
We shall henceforth assume the shell to be
static in the sense of general relativity,
i.e., there exists a time-like,
hypersurface-orthogonal Killing vector field of the metric within
the shell, which Killing vector is also a symmetry of the
shell's stress-energy tensor. But although the metric inside
the shell is flat (i.e., its Riemann tensor vanishes),
in the case of general relativity there
is no {\it unique} way to physically associate a Minkowski metric with the
points of the interior. As a solution to the field equations,
the hole argument applies just as well to the Minkowski metric
as to any other empty-space solution
(see, e.g., Refs.~\cite{stachelY,johnextra}).
But the presence of non-gravitational physical processes, previously
regarded as presenting a difficulty, because of their gravitational effects,
now proves to be an
advantage: They enable us to define physically the spatial points and
temporal instants associated with physical events
inside the shell by relating them to events on the shell.
To see how this can be done, let us look first at the situation in two-dimensional
space-time. Here the sphere reduces to two parallel material strips,
representing the histories of the two sides of the ``one-sphere.''
In the inertial frame fixed by these strips, let the distance between the inner
surfaces of the two sides of the ``one-sphere" be 2R. Let an event $e$ in the interior
be marked by the
emission of two light rays, one travelling towards each of
the two sides.
Assuming that times on the two strips have been
synchronized (even in general relativity, a unique global time exists
in the case of static metrics),
we easily see that the time and position of the event, $\tau_e$
and $R_e$ respectively, are given by (see Figure 4)
\begin{equation}
\tau_e = {\tau_{eR1} + \tau_{eR2} \over 2} - R ~,~~~
R_e = {\tau_{eR1} - \tau_{eR2} \over 2} ~.
\label{sphere1}
\end{equation}

\begin{figure}
\centering
\includegraphics[height=120mm]{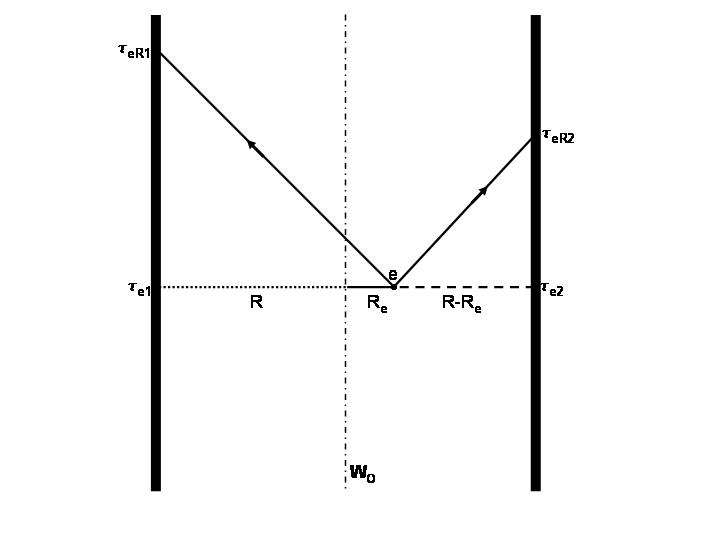}
\caption{In two-dimensional space-time, a ``spherical shell"  reduces to two strips. We denote
by $R$ the ``radius" (the half of the distance between the strips) and by $R_e$
the position of the event $e$ with respect to the origin of the position axis, which coincides with
the ``center" of the spherical shell. In figure $W_O$ is the world-line of this origin of the position
axis.}
\label{fig4}
\end{figure}

This procedure can be generalized to higher dimensions as follows.
If we increase the number of dimensions by rotating one of the strips of matter
(the other strip is not needed; cf. our discussion in Section 2
of the transition from two to three dimensions), then the strip
of matter becomes a space-time ``cylinder of matter", {\it i.e.} the history in
time of a circle (or more generally a spherical annulus for three spatial
dimensions, etc).
Again, we shall assume that clocks are distributed
over the surface of this cylinder at known positions and synchronized,
enabling the recording of the arrival time
of any light signal from events inside the cylinder.

The light ray emitted from $e$ now becomes a retarded light cone
(of the appropriate number
of dimensions). That is, we must now postulate that the event $e$ includes the
emission of a (retarded) light cone, which intersects the material cylinder
in some curve. Let the times of arrival at the material cylinder
of the light rays on this cone
at the material
cylinder be noted by synchronized clocks. Of course,
the fixed places of these clocks are known.

We shall now show that our previous 2D analysis
can be applied to this situation.
First of all, note that, if the event $e$ does not lie on the central axis
of the cylinder (the case in which it does is trivial: all the signals of
the light cone will arrive at the same time), then the central axis and $e$
define a time-like two-plane. If we are able to pick out the signals of the light cone
that lie in this two-plane, then we have reduced the problem to the two-dimensional
one. But, as a moment's thought shows, of all the times recording the arrival of
the light-cone signals, the two that lie in this plane will be the earliest
and the latest recorded, respectively. Taking the 3D case, for example,
and looking at the circular spatial cross-section of the cylinder at any time,
one sees that points on the circle not at the ends of the diameter through the
position of $e$ are further away than the closer point of the diameter.

Thus one need merely determine the latest and the earliest times of arrival
on the cylinder of signals
from the event.
Calling them $\tau_{eR1}$ and $\tau_{eR2}$,
respectively, we can again apply Eq.~(\ref{sphere1}) to find the time and
position of $e$. Of course, Eq.~(\ref{sphere1}) only gives us the magnitude
of $R_e$, but the positions of the clocks
recording the earliest and the latest times already fix the diameter along
which $e$ lies.

There may be other and better ways of fixing the position and time of
the event $e$,
but at least we have demonstrated that one such procedure exists.
It is also interesting to note that, since it uses the entire
light cone, this method does not seem to require the measurement of an angle.

Having seen how to define the spatial point and
temporal instant associated with an event inside the shell by
relating them to events on the shell, we no longer need the
Wigner-Salecker microscopic interior clock. As indicated above, to make a
microscopic clock reading meaningful, we need to relate it
to events on the shell, and we have shown how to do this directly
for the events under study.

This might also have interesting implications
for the problem of quantum limitations in the gravitational
context, which is transformed into the problem of what limitations
are imposed by the quantum of action on the ability to define
the position and time of an event inside the shell in terms of
measurements of positions and times by macroscopic clocks on the shell.

\section{Summary and outlook}
The analysis in Sections~3-5 generalizes the Wigner-Salecker
procedure in such a way that it is now applicable
to the measurement of arbitrary intervals in an arbitrary number
of space-time dimensions.
A later paper will take up a detailed analysis of quantum limits to
their measurability, but some of the results presented here
are already relevant to the ongoing debate on ``Planck-scale
uncertainty principles".
Several studies~\cite{ng1994,gacmpla,adlerSW}
have proposed such
uncertainty principles on the basis of the 2D Wigner-Salecker
procedure for the measurement of space-like intervals.
The key aspect of those studies is the role played by (advanced and
retarded) proper time measurements.
Our generalization to the measurement of arbitrary intervals
renders straightforward a corresponding generalization of the analyses
reported in Refs.~\cite{ng1994,gacmpla,adlerSW} to the case of non-space-like intervals.

Extending this procedure to more than two space-time dimensions,
the measurement of
advanced and retarded times is  still a key ingredient; but
an angle measurement also may be needed. If so, this result suggests
a clear path for generalization of these 2D
analyses to the case of
higher space-time dimensions. The search for
possible quantum (-gravity) limitations
on the accuracy of this angular measurement would then be a key issue
for such a generalization\footnote{Our classical angular
measurement analysis can provide the basis for a corresponding quantum-mechanical
analysis, but presumably, since
the studies reported
in Refs.~\cite{ng1994,gacmpla,adlerSW} do not advocate exactly the
same perspective on quantum-measurability analysis of the Wigner-Salecker procedure
already in the 2D case, different authors might treat differently the
angular analysis at the quantum level. In a future paper, we shall propose
our own perspective.}.

In Section~6 we propose a modification of the
Wigner-Salecker
procedure using a ``spherical shell setup", and in Section~7 show how it may be extended
to include gravitation,
 which should allow a more consistent and theoretically comprehensive
derivation of measurability limits relevant to quantum-gravity research.
This
will of course require a detailed and, in large part, new
quantum measurability analysis: It will probably be necessary
to treat the time measurements on the shell
in close analogy with what has been done for such measurements
by a Wigner-Salecker {\it microscopic} clock, but some new
elements will arise from the requirement that the
system of clocks on the shell be {\it macroscopic}, having a fixed position
and capable of recording the time of arrival of signals.
While in classical general relativity there is no in-principle
obstruction to rigid motions, in a quantum setting the shell of finite
mass and the clocks included in it are subject to the
Heisenberg indeterminacy principle
and can be set in rigid motion (rather than forming a ``rigid body")
only up to a certain accuracy, which must be established.

Another important point to be considered for future work
concerns the problem of developing a measurement procedure
for the physical components of the space-time curvature, and
showing how the existence of the quantum of action leads to
 a ``Planck-scale
indeterminacy principle" for the curvature components,
as already carried out in linear approximation
in Ref.~\cite{bergstac}. As argued in Ref.~\cite{wign},
this should be possible with a suitable modification of
the Wigner-Salecker procedure,
even though in its original form,
and in our generalization,
the procedure assumes the absence of curvature in the
space-time region of interest for the measurement procedure.

\bigskip
\bigskip
\bigskip
\bigskip
We thank Andor Frenkel for valuable
comments on an earlier draft.
Parts of this paper were presented by one of us (J.S.)
at the Center for Gravitational Physics and Geometry of
Pennsylvania State University. He
thanks Abhay Ashtekar for his hospitality.

\baselineskip 12pt plus .5pt minus .5pt

\end{document}